\documentclass[onecolumn,aps,prc,superscriptaddress]{revtex4}
\usepackage{epsfig,graphics}
\usepackage{graphicx}% Include figure files
\usepackage{dcolumn}% Align table columns on decimal point
\usepackage{bm}% bold math
\usepackage{amsmath}
\usepackage[usenames]{color}
\usepackage{ulem} %% for strike-through
\usepackage[colorlinks,linkcolor=blue,urlcolor=blue,citecolor=blue]{hyperref}
\begin{document}

\title{Thermodynamics of partonic matter in relativistic heavy-ion collisions from a multiphase transport model}

\author{Han-Sheng Wang}
\affiliation{Key Laboratory of Nuclear Physics and Ion-beam Application (MOE), Institute of Modern Physics, Fudan University, Shanghai 200433, China}
\affiliation{Shanghai Research Center for Theoretical Nuclear Physics, NSFC and Fudan University, Shanghai $200438$, China}
\author{Guo-Liang Ma}
\email[]{glma@fudan.edu.cn}
\affiliation{Key Laboratory of Nuclear Physics and Ion-beam Application (MOE), Institute of Modern Physics, Fudan University, Shanghai 200433, China}
\affiliation{Shanghai Research Center for Theoretical Nuclear Physics, NSFC and Fudan University, Shanghai $200438$, China}
\author{Zi-Wei Lin}
\email[]{LINZ@ecu.edu}
\affiliation{Department of Physics, East Carolina University, C-209 Howell Science Complex, Greenville, NC 27858}
\author{Wei-jie Fu}
\email[]{wjfu@dlut.edu.cn}
\affiliation{School of Physics, Dalian University of Technology, Dalian, 116024, China}

%\date{\today}

\begin{abstract}
Using the string melting version of a multiphase transport model, we focus on the evolution of thermodynamic properties of the central cell of parton matter produced in Au$+$Au collisions ranging from 200 GeV down to 2.7 GeV. The temperature and chemical potentials have been calculated based on both Boltzmann and quantum statistics in order to locate their evolution trajectories in the QCD phase diagram. We demonstrate that the trajectories can depend on many physical factors, especially the finite nuclear thickness at lower energies. However, from the evolution of pressure anisotropy, only partial thermalization can be achieved when the partonic systems reach the predicted QCD phase boundary. It provides some helpful insights to studying the QCD phase structure through relativistic heavy-ion collisions. 

\end{abstract}


\maketitle

\section{Introduction}
\label{introduction}

The ultrarelativistic heavy-ion collisions at the BNL Relativistic Heavy Ion Collider (RHIC) and the CERN Large Hadron Collider (LHC) have created a partonic matter at extreme conditions of temperature and energy densities, the quark-gluon plasma (QGP), which is governed by the quantum chromodynamics (QCD) theory. The first-principles lattice QCD calculation shows that the transition from hadronic to the partonic matter at zero baryon chemical potential $\mu_B$ is a smooth crossover \cite{Aoki:2006we,Bellwied:2015rza,Bazavov:2018mes}. But the calculation of phase transition in the QCD phase diagram at finite baryon chemical potential still has large uncertainties~\cite{Fischer:2014ata,Gao:2015kea,Fu:2019hdw}, especially regarding the conjectured endpoint of the first-order phase transition boundary that is the so-called QCD critical endpoint (CEP) \cite{Gavai:2008zr,Stephanov:2004wx,Mohanty:2009vb}, due to the famous sign problem \cite{Gavai:2003mf,deForcrand:2002hgr,Fodor:2001au}.

To explore the nature of the QCD phase diagram, the beam energy scan (BES) program at RHIC is searching for the QCD critical point with Au$+$Au collisions at a large range of collision energies~\cite{Aggarwal:2010wy,Adamczyk:2013dal,Adamczyk:2014fia,Adare:2015aqk,Adamczyk:2017wsl,Adam:2020unf}. The fireballs created in Au$+$Au collisions at different energies freeze-out at different points of the QCD phase diagram. Because certain singularities will appear at the CEP in the thermodynamic limit~\cite{ibook02}, we expect to observe certain nonmonotonic behaviors if the evolution trajectory of the colliding system is close enough to the CEP. For example, event-by-event fluctuations of various conserved quantities are proposed as possible signatures of the existence of the CEP \cite{Koch:2005vg,Asakawa:2000wh,Asakawa:2009aj} because they are proportional to the corresponding susceptibilities and correlation lengths. Many recent experimental results on net-proton fluctuations hint that a critical point might have been reached during the evolution of Au$+$Au collisions at a low collision energy~\cite{Adamczyk:2013dal, Adam:2020unf,Luo:2017faz}, which serves as a main motivation for the upcoming research projects such as those at FAIR in Germany, NICA in Russia, and HIAF in China.

On the other hand, it is difficult to connect thermal properties of static QCD matter with the experimental measurements, since relativistic heavy-ion collisions involve different dynamical evolution stages. To study the full evolution history of the thermodynamic properties of the QCD matter with a dynamical transport model may serve as a bridge between the gap~\cite{Zhang:2008zzk, Lin:2014tya}. In this work, we investigate the space-time evolution of the parton matter created in Au$+$Au collisions at different energies, including transverse flow, effective temperature and conserved charge chemical potential by using the string melting version of a multiphase transport (AMPT) model~\cite{Lin:2004en}.

The paper is organized as follows. Section~\ref{AMPT} briefly introduces the string melting version of AMPT model and the improvements that we make. Comparison of the space-time evolution of transverse flow at different collision energies are presented in Sec.~\ref{flow}. We then discuss the space-time evolution of the effective temperature and chemical potentials in Sec.~\ref{SPACE-TIME EVOLUTION}. We show the trajectories of Au$+$Au collisions at different energies in the QCD phase diagram in Sec.~\ref{diagram}. We present the space-time evolution of pressure anisotropy to discuss the systems are in equilibrium or nonequilibrium in Sec.~\ref{Pressureanisotropy}. Finally, a summary is given in Sec.~\ref{summary}.

\section{A multiphase transport model including the nuclear thickness}
\label{AMPT}

The string melting version of the AMPT model consists of fluctuating initial conditions from the heavy-ion jet interaction generator (HIJING) model~\cite{Wang:1991hta}. In this model, minijet partons and strings are produced from hard processes and soft processes, respectively. With the string melting mechanism, all parent hadrons from the fragmentation of the excited strings are converted into partons. The interactions among these partons are described by Zhang's parton cascade (ZPC) model \cite{Zhang:1997ej}, which includes elastic two-body scatterings based on the leading order pQCD gg $\rightarrow $ gg cross section:
\begin{equation}
\frac{d\sigma}{dt}=\frac{9\pi\alpha^{2}_{s}}{2}(1+\frac{\mu^{2}}{s})\frac{1}{(t-\mu^{2})^{2}}.
\label{q1}
\end{equation}
In the above, $\alpha_{s}$ is the strong-coupling constant (taken as 0.33), while $s$ and $t$ are the usual Mandelstam variables. The effective screening mass $\mu$ is taken as a parameter in ZPC for the parton scattering cross section, and we set $\mu$ as 2.265 fm$^{-1}$ leading to a total cross section of about 3 mb for elastic scatterings in the default setting. The AMPT model implements a spatial quark coalescence model, which combines nearby freeze-out partons into mesons or baryons, to describe the transition from the partonic matter to the hadronic matter. The final-stage hadronic evolutions are modeled by an extension of a relativistic transport model (ART) including both elastic and inelastic scatterings for baryon-baryon, baryon-meson and meson-meson interactions~\cite{Li:1995pra}. Our other parameters are taken as same as those from Ref.~\cite{Lin:2014tya,Ma:2016fve}, which can reasonably reproduce many experimental observables such as rapidity distributions, $p_T$ spectra, and anisotropic flows \cite{Lin:2001zk,Chen:2004dv,Ma:2016fve} for both Au$+$Au collisions at RHIC and Pb$+$Pb collisions at LHC energies.

\begin{figure}[htbp]
	\centering\includegraphics[scale=0.4]{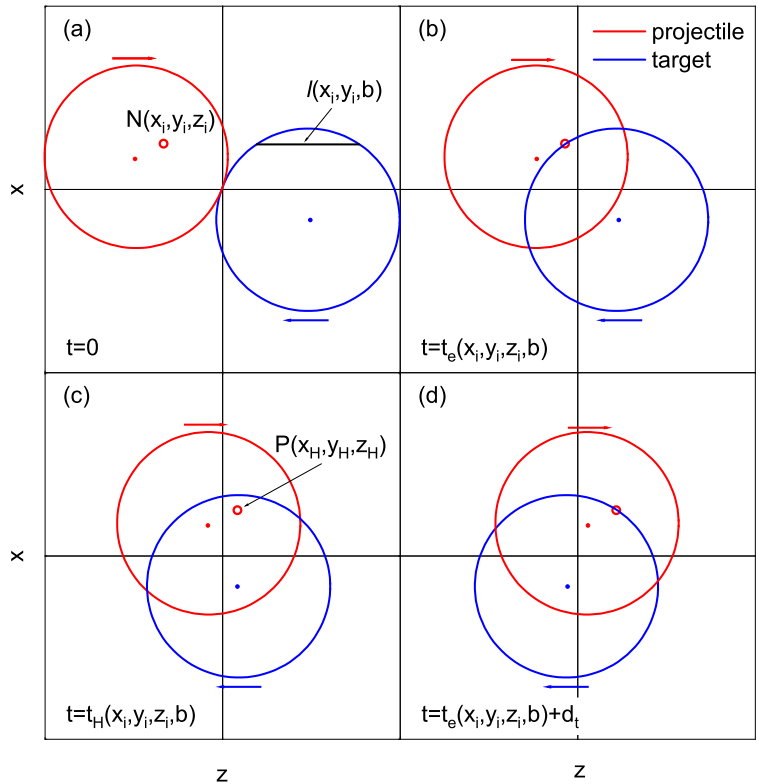}
        \caption{(Color	online) The schematic diagrams of a Au$+$Au collision with an impact parameter $b$ in the $x$-$z$ plane. (a) Consider a projectile nucleon $N$ (small open circle) at a location of ($x_i, y_i, z_i$) at the starting time $t=0$; (b) the projectile nucleon enters the target nucleus at $t=t_e(x_i, y_i, z_i, b)$; (c) the wounded nucleon from the projectile produces parent hadrons at a location of ($x_H, y_H, z_H$) at $t=t_H(x_i, y_i, z_i, b)$; (d) the projectile nucleon leaves the target nucleus at $t=t_e(x_i, y_i, z_i, b)+d_t$.}
\label{schematic_diagram}
\end{figure}

To study heavy-ion collisions at low energies, we have improved the string melting AMPT by modeling the finite nuclear thickness, which has been shown to be important for nuclear collisions at lower energies \cite{Lin:2017lcj, Mendenhall:2020fil, Mendenhall:2021maf}. In our convention, the $x$ axis is chosen along the direction of impact parameter $b$ from the target center to the projectile center, the $z$ axis is along the projectile direction, and the $y$ axis is perpendicular to both the $x$ and $z$ directions.
We consider the moment when the projectile and target nuclei contact each other as the starting time $t=0$, while the proper time $\tau$ is defined as $(t^2-z^2)^{1/2}$. The spatial density of nucleons inside projectile or target follows the Woods-Saxon distribution. As shown in Fig.~\ref{schematic_diagram}(a), for a nucleon inside a hard-sphere projectile located at an initial position of ($x_i, y_i, z_i$), the thickness length $l$ of target that the projectile nucleon punches through can be calculated as follows,
\begin{eqnarray}
	l(x_i,y_i,b)=2\sqrt{R^2-(x_i\pm b/2)^2-y_i^2},
	\label{thickness}
\end{eqnarray}
where $R$ is the hard-sphere radius of colliding nuclei, and $\pm$ applies to projectile or target nucleons, respectively. 
As shown in Fig.~\ref{schematic_diagram}(b), the time $t_e$ when the projectile nucleon enters the target in the center-of-mass frame of a Au$+$Au collision can be calculated as follows,
\begin{eqnarray}
%	t_e(x_i,y_i,z_i,b)&=&\frac{\sqrt{R^2-b^2/4}-\lift[l(x_i,y_i,b)/2 \pm z_i\rigit]}{2\mathrm{sinh}~y_{CM}}, 
         t_e(x_i,y_i,z_i,b)&=&\frac{\sqrt{R^2-b^2/4}-[l(x_i,y_i,b)/2 \pm z_i]}{2\mathrm{sinh}~y_{CM}}, 
	\label{teq}
\end{eqnarray}
where $y_{CM}$ is the projectile rapidity in the center-of-mass frame.

Since parent hadrons are produced by interactions between projectile and target nucleons, as shown in Fig.~\ref{schematic_diagram}(c), the production time of parent hadrons, $t_H$, is obtained by sampling according to a time profile based on the probability function ~\cite{Lin:2017lcj},
\begin{equation}
	\frac{d^2E_T}{dy dt_H}=a_n [(t_H-t_e)(t_e+d_t-t_H)]^n \frac{dE_T}{dy}, t_H\in [t_e, t_e+d_t],
	\label{time_profile}
\end{equation}
where we take the power as $n = 4$, $a_n = 1/d_t^{2n+1}/\beta(n+1, n+1)$ is the normalization factor with the $\beta$ function of $\beta(a, b)$, and $d_t=l/(2\mathrm{sinh}~y_{CM})$ is the duration time during which the projectile nucleon completely crosses the target nucleus. The parent hadrons produced by same projectile or target nucleon are assumed to be produced at the same time of $t_H$. Then the longitudinal coordinate of a parent hadron can be obtained as follows:
\begin{equation}
	z_H=z_i\pm t_H\mathrm{sinh}~y_{CM},
	\label{zstring}
\end{equation}
while its transverse coordinates ($x_H, y_H$) are set to the transverse positions of the projectile or target nucleon.

In the following, the partons are generated by string melting after a formation time:
\begin{equation}
	t_f=E_{H}/m^2_{T,H},
	\label{tf }
\end{equation}
where $E_{H}$ and $m_{T,H}$ represent the energy and transverse mass of the parent hadron. The initial positions of partons from melted strings are calculated from those of their parent hadrons using straight line trajectories. As a result, the initial condition of partonic matter after considering the finite-thickness effect is used for the parton cascade simulations in this study. 

To study the thermodynamics properties of partonic matter, we focus on the space-time evolution of partonic matter during the process of parton cascade only in this study. Using the string-melting version of the AMPT model with the finite-thickness effect, 10 000 events of Au$+$Au central collisions ($0 - 5\%$ centrality modeled with b$\leq$3 fm) are generated for each energy ($\sqrt {s_{NN}}$ = 200, 62.4, 39, 27, 19.6, 11.5, 7.7, 4.9, and 2.7 GeV) which can be provided by the RHIC, FAIR, and NICA facilities. 

\section{Results and discussions}
 \label{results}

 \subsection{Space-time evolution of transverse flow}
 \label{flow}
 
\begin{figure}[htbp]
	\centering\includegraphics[scale=0.4]{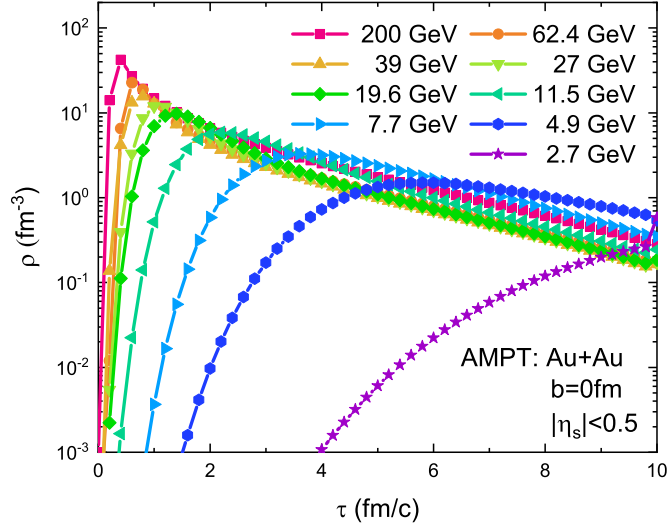}
        \caption{(Color	online) Proper-time evolution of parton density averaged over the transverse area of the overlap volume within space-time rapidity $|\eta_s|<0.5$ in $b=0$ fm Au$+$Au collisions at different energies.}
\label{denx3}
\end{figure}

First, the densities of formed partons averaged over the transverse area of the overlap volume within space-time rapidity $|\eta_s|<0.5$ as functions of proper time in $b=0$ fm Au$+$Au collisions at different energies are shown in Fig.~\ref{denx3}. The nuclear transverse area $A_T$ \cite{Lin:2017lcj} is defined as:
\begin{eqnarray}\label{transverse}
A_T=
\begin{cases}
\pi R^2_A & {t\ge d_t^{nuclei}/2}\\
\pi R^2_A \left [ 1-(1-2t/d_t^{nuclei})^2\right ] & {t<d_t^{nuclei}/2}
\end{cases},
\end{eqnarray}
with $R_A=1.12A^{1/3}$ fm, $A=197$, and $d_t^{nuclei} = 2R_A/\mathrm{sinh}~y_{CM}$ is the duration time for two nuclei of the same mass number $A$ with $b=0$ fm to cross each other in the center-of-mass frame. The density increases with the proper time at first, because more partons are produced. Higher density is reached at higher collision energy. With the expansion of the fireball, the density decreases gradually. Both the increase and the decrease become slower at lower collision energies, since the nuclei have a larger thickness at lower collision energies which slows down the evolution, especially in the longitudinal direction.  

\begin{figure*}[htbp]
	\centering
	\includegraphics[width=1\linewidth]{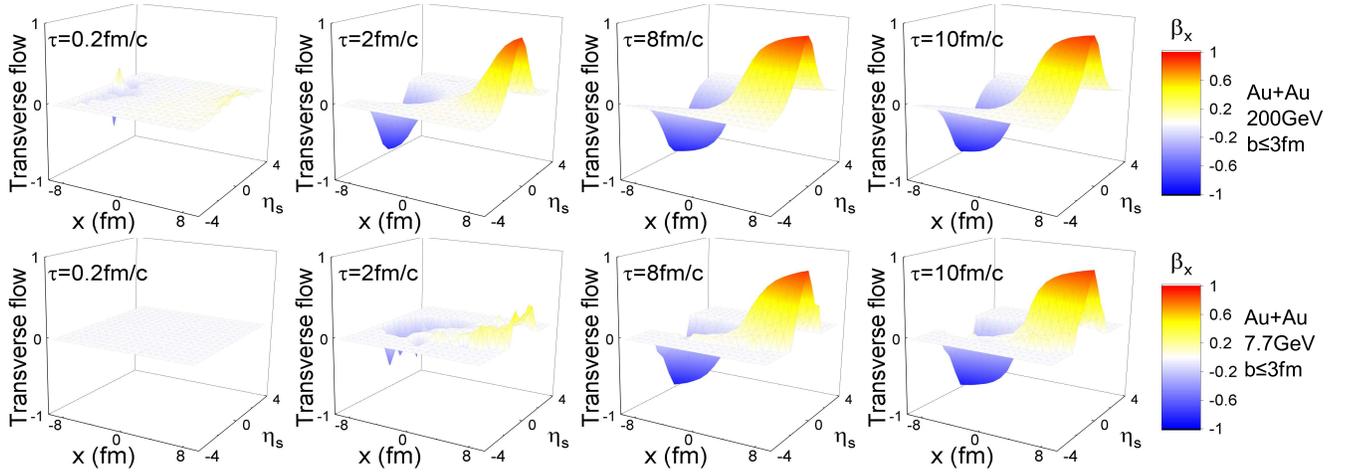}
	\caption{(Color	online) Transverse flow component $\beta_x$ along the $x$ axis ($|y| < 0.5$ fm) as functions of $x$ and $\eta_s$ at different proper times in central Au$+$Au collisions at $\sqrt{s_{NN}}$ = 200 GeV (first row) and 7.7 GeV (second row).}
	\label{Tr_flow}
\end{figure*}

At the same time, the radial flow is calculated employing $\vec{\beta}=(\sum_{i}\vec{p}_i/\sum_{i}E_i)$, where the sum over index $i$ takes into account all partons in the cell for all events of a given collision system.  Flow component along the $x$ direction $\beta_x$ as functions of coordinate $x$ and space-time rapidity $\eta_s$ at different times in cells within $|y| < 0.5$ fm in central Au$+$Au collisions at $\sqrt{s_{NN}}$ = 200 GeV and 7.7 GeV are shown in Fig.~\ref{Tr_flow}.  We can see the antisymmetry of the transverse flow along the $x$ axis in space-time rapidity, after averaging over many events of central collisions. The flow is very small at the early time $\tau = 0.2$ fm/$c$ and then develops rather faster, especially at larger $x$~\cite{Lin:2014tya}. 

\begin{figure}[htbp]
	\centering\includegraphics[scale=0.4]{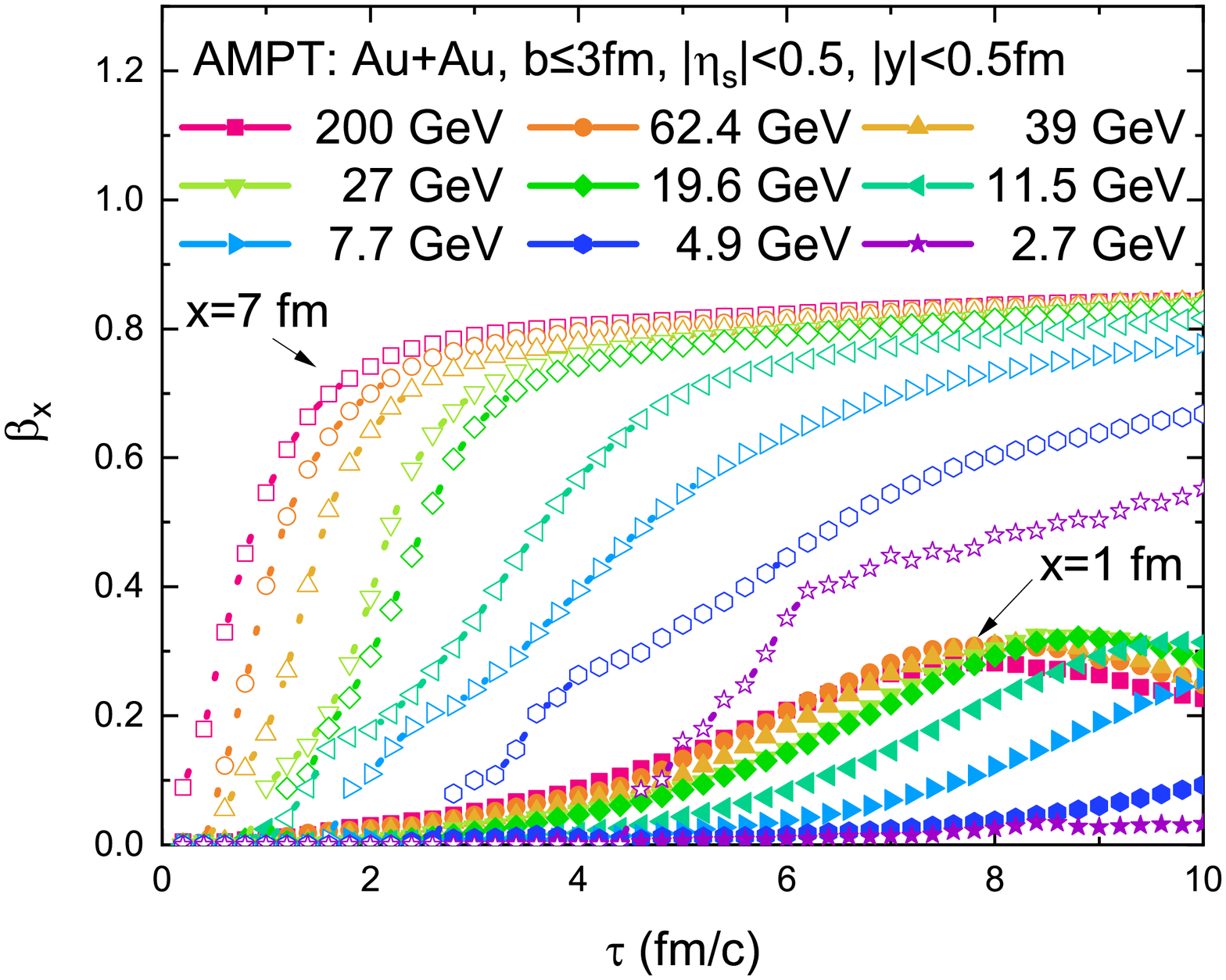}
	\caption{(Color	online) Proper time evolution of transverse flow component $\beta_x$ of partons within space-time rapidity $|\eta_s|<0.5$ in the cells at ($x, y$)=(1 fm, 0 fm) (filled symbols) and  ($x, y$)=(7 fm, 0 fm) (open symbols) in central Au$+$Au collisions at different energies.}
	\label{flow_t}
\end{figure}

Figure~\ref{flow_t} shows the transverse flows of partons in the two selected cells at ($x, y$)=(1 fm, 0 fm) and ($x, y$)=(7 fm, 0 fm) within space-time rapidity $|\eta_s|<0.5$ as functions of proper time in central Au$+$Au collisions at different energies. The transverse flow is bigger further away from the center of the overlap volume of central collisions \cite{Lin:2014tya}. We see that the transverse flow increases with time for both the inner cell and the outer cell in the beginning. Because the parton density increases faster at higher collision energy, the transverse flow grows faster at higher collision energy for both the inner and outer cells. However, compared with the case of parton density in Fig.~\ref{denx3}, the development of transverse flow generally shows a time delay is slower.

\subsection{Space-time evolution of temperature and chemical potentials}
\label{SPACE-TIME EVOLUTION}

In the AMPT model, the energy-momentum tensor, $T^{\mu \nu }$ can be calculated by
averaging over particles and events in a volume $V$~\cite{Zhang:2008zzk}, i.e.,
\begin{equation}
T^{\mu \nu }=\frac{1}{V}\sum_{i}\frac{p_{i}^{\mu }p_{i}^{\nu
}}{E_{i}}.
\end{equation}
In the rest frame of a small volume cell, the energy density can be given by $\epsilon = T^{00}$, while the pressure components are related to the energy-momentum tensor by $P_{x} = T^{11}$, $P_{y} = T^{22}$, $P_{z} = T^{33}$. The net conserved charge number densities $n_{B}$, $n_{Q}$, and $n_{S}$ can be calculated for the given volume as well. Therefore, the corresponding chemical potentials $n_{B}$, $n_{Q}$, and $n_{S}$, and $T$ can be obtained by numerical solving Eqs.~(\ref{mu_T}) after the net conserved charge densities $n_{B}$, $n_{Q}$, and $n_{S}$ and $\epsilon$ are obtained through the AMPT model. Note that in this study we only extract $\mu$ and $T$ values for the center cell, 
for which the rest frame is assumed to be A$+$A collision center-of-mass frame. 

\begin{figure}[htbp]
	\includegraphics[scale=0.37]{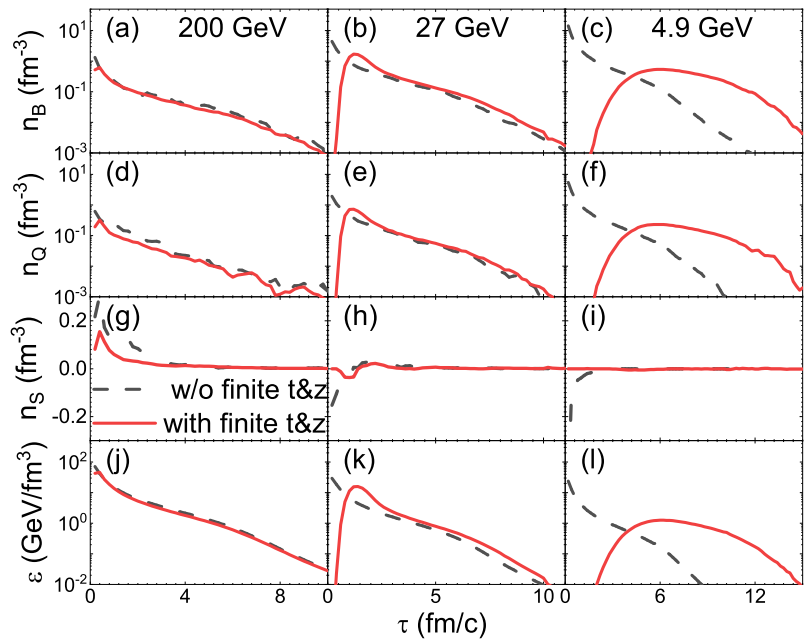}
	\caption{(Color	online) Proper-time evolution of net baryon number density $n_{B}$ (first row), net electric charge density $n_{Q}$ (second row), net strangeness number density $n_{S}$ (third row), and energy density $\epsilon$ (fourth row) for the central cell in central Au$+$Au collisions at 200 GeV (left column), 27 GeV (middle column), and 4.9 GeV (right column) with (solid) and without (dashed) including the finite nuclear thickness.}\label{nB_epsilon}
\end{figure}

Figure~\ref{nB_epsilon} shows the proper-time evolution of net baryon number density $n_{B}$, net electric charge density $n_{Q}$, net strangeness number density $n_{S}$, and energy density $\epsilon$ for the central cell, defined as the cell within ($|x| < 0.5$ fm, $|y| < 0.5$ fm) and the space-time rapidity range of $|\eta_s| < 0.5$, in central Au$+$Au collisions at three selected beam energies from the AMPT-SM model. At the top RHIC energy of 200 GeV, the results with and without the finite nuclear thickness are almost the same \cite{Lin:2017lcj, Mendenhall:2020fil}. With the decrease of the beam energy, the peak energy and charge densities are reached later due to the longer time that two nuclei take to cross each other. Therefore, it is important to consider the finite nuclear thickness effect for simulating heavy-ion collisions at low beam energies \cite{Lin:2017lcj, Mendenhall:2020fil, Mendenhall:2021maf}. Note that we show the results with the finite nuclear thickness effect in the rest of this paper, unless stated otherwise. In addition, we see that the net strangeness number density can be negative at low energies in the central cell. This is because of the large baryon density, which leads to most $s$ in $\Lambda$ but most $\Bar{s}$ in K. Since the quark formation time is inversely proportional to the parent hadron transverse mass in AMPT's string melting, $s$ from $\Lambda$ has a smaller formation time than $\Bar{s}$, which produces negative $n_S$ at early times.

\begin{figure}[htbp]
	\includegraphics[width=1\linewidth]{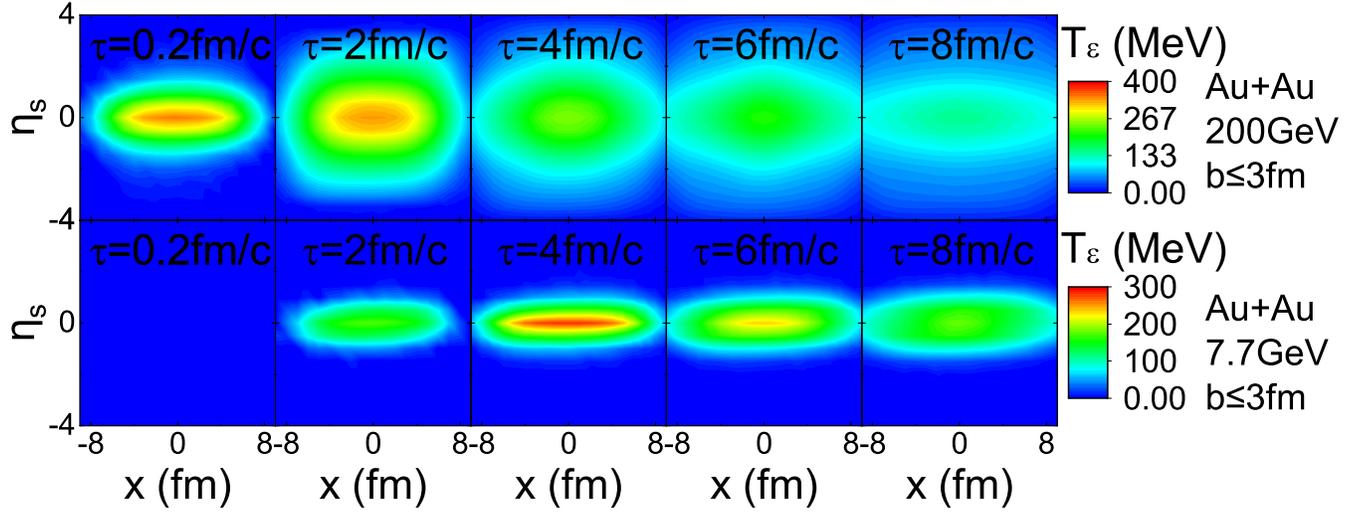}
	\caption{(Color	online) Contour plots of the effective temperature from Boltzmann statistics as a function of the $x$ coordinate and space-time rapidity $\eta_{s}$ at different proper times in central Au$+$Au collisions at $\sqrt{s_{NN}}$ = 200 GeV (top row) and 7.7 GeV (bottom row) for the parton matter within $|y|<$0.5 fm.}\label{contours}
\end{figure}

The two-dimensional (2D) distributions of extracted local temperature from
Boltzmann statistics as functions of coordinate $x$ and space-time rapidity at different proper times in central Au$+$Au collisions at $\sqrt{s_{NN}}$ = 200 and 7.7 GeV are shown in Fig.~\ref{contours}. We can see that the highest temperature is reached at the center of the overlap region after the two nuclei overlap completely ($\tau \approx $ 0.2 and 4 fm/$c$ for 200 and 7.7 GeV, respectively). After that moment, the temperature decreases with the evolution of the expanding system.

\begin{figure}[htbp]
	\centering\includegraphics[scale=0.4]{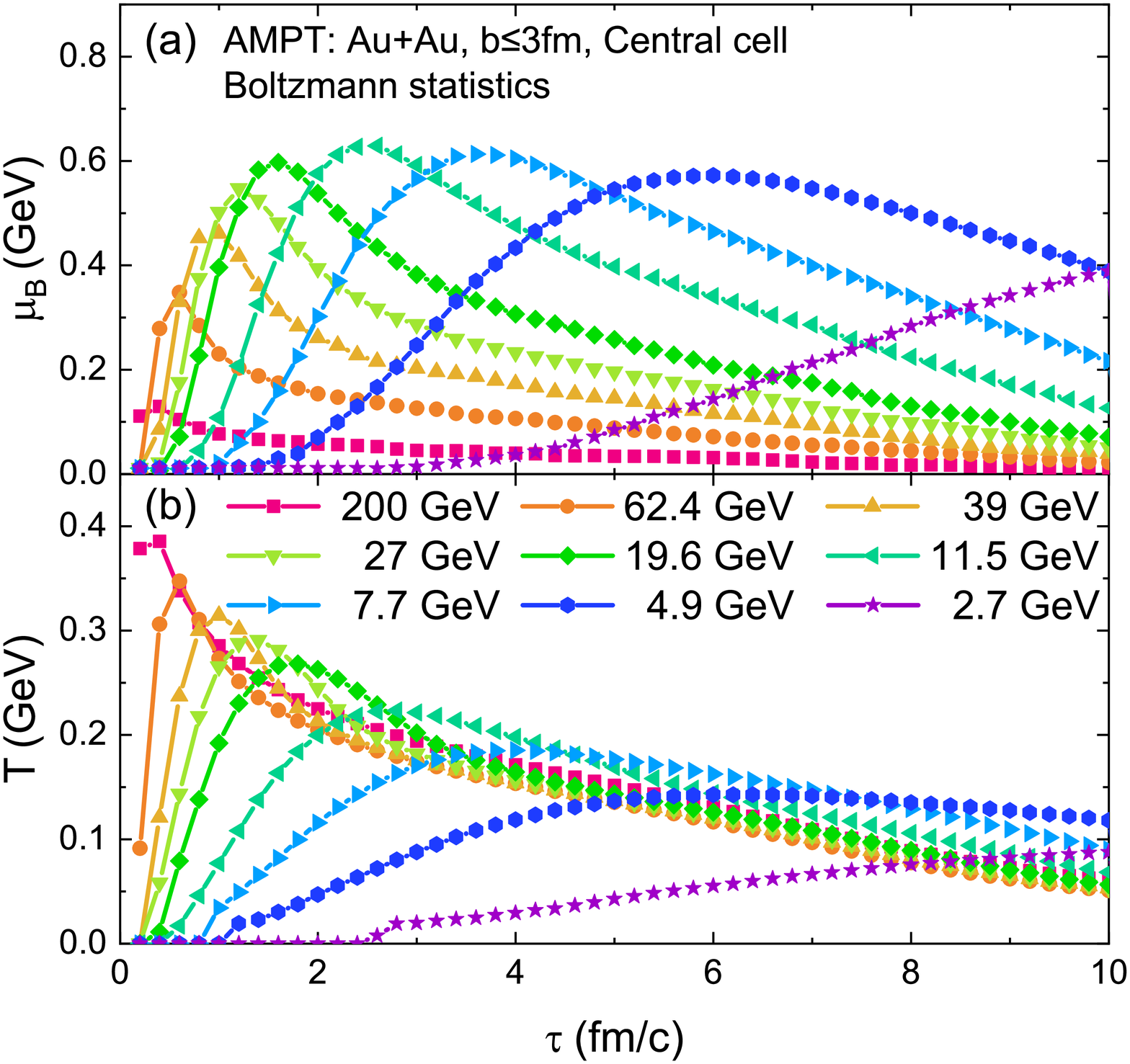}
	\caption{(Color	online) Proper time evolution of (a) baryon chemical potential $\mu_{B}$ and (b) temperature $T$ for the central cell in central Au$+$Au collisions at different energies.}
	\label{T_mub}
\end{figure}

The proper-time evolutions of the baryon chemical potential $\mu_{B}$ and temperature $T$ for the central cell in central Au$+$Au collisions at different beam energies from the AMPT-SM model are shown in Fig.~\ref{T_mub}(a) and \ref{T_mub}(b), respectively. We can see that both baryon chemical potential and temperature increase with time at first, then they decrease with time, which indicates that the collision system is first compressed and heated, and then becomes dilute and cools down due to the expansion. However, the energy dependencies of the baryon chemical potential and temperature are different. Figure.~\ref{T_mub}(b) shows that a higher temperature is reached at a higher collision energy; in contrast, the highest baryon chemical potential is achieved at an intermediate energy of $\sqrt{s_{NN}}$ = 7.7 GeV, as shown in Fig.~\ref{T_mub}(a). In general, the time evolution at lower energies is slower than that at higher energies due to the influence of the finite nuclear thickness.

\begin{figure}[htbp]
	\centering\includegraphics[scale=0.4]{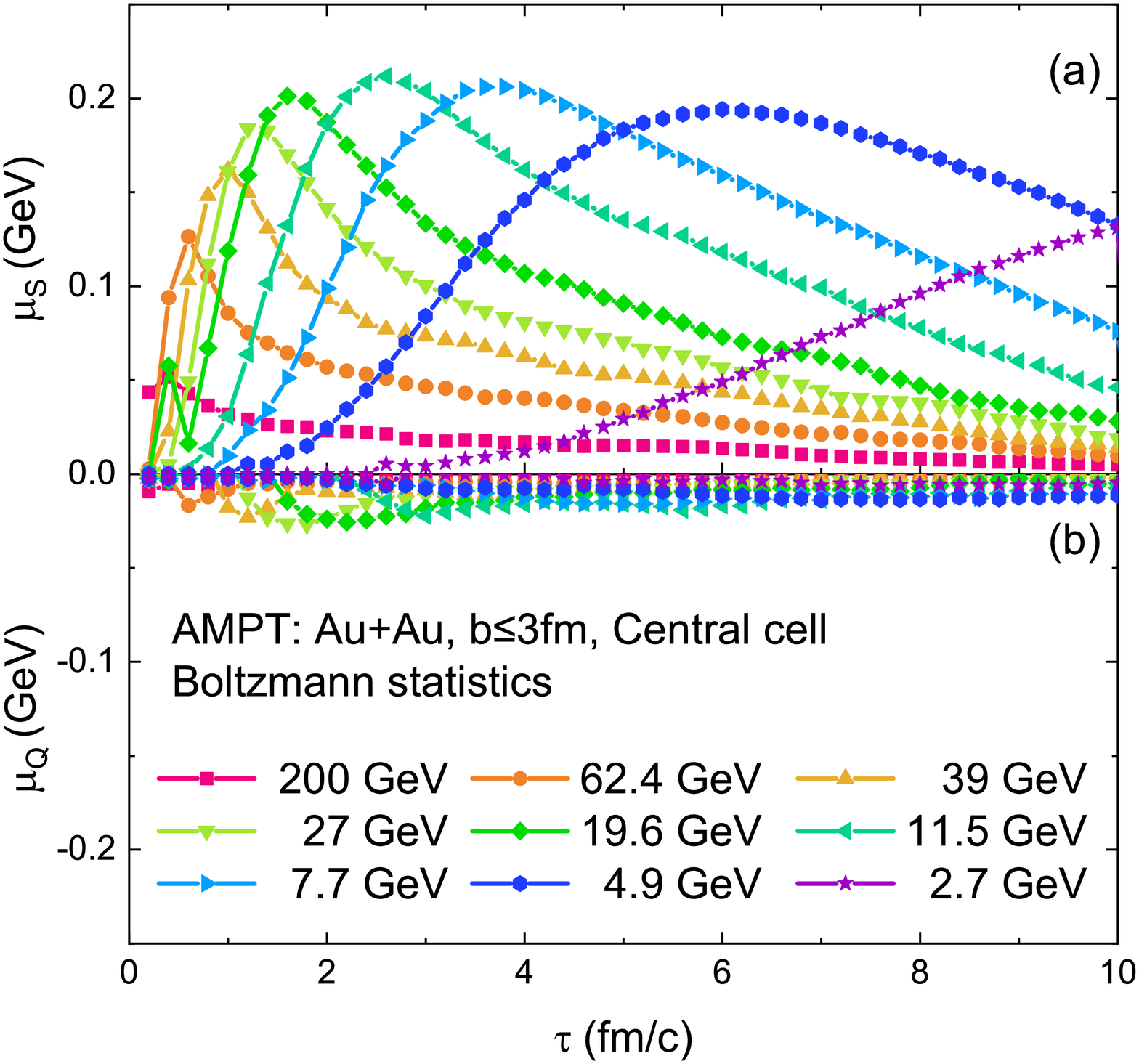}
	\caption{(Color	online) Proper-time evolution of (a) chemical potentials of  strangeness $\mu_{S}$ and (b) electric charge $\mu_{Q}$ for the central cell in central Au$+$Au collisions at different energies.}
	\label{muq_nus}
\end{figure}
The proper-time evolutions of the chemical potentials of electric charge $\mu_{Q}$ and strangeness $\mu_{S}$ for the central cell in central Au$+$Au collisions at different beam energies from the AMPT-SM model are shown in Fig.~\ref{muq_nus}(a) and \ref{muq_nus}(b), respectively. We obtain positive $\mu_{S}$ but negative $\mu_{Q}$. The $\mu_{S}$ is seen to be roughly proportional to $\mu_{B}$, i.e., $\mu_S \approx 1/3\mu_B$, while the magnitude of $\mu_Q$ is very small. We observe that the magnitudes of two chemical potentials increase with time at first, and then decrease with time, which follow a similar trend as $\mu_B$. 

\subsection{Trajectories in the QCD phase diagram}
\label{diagram}

\begin{figure}[htbp]
	\centering\includegraphics[scale=0.4]{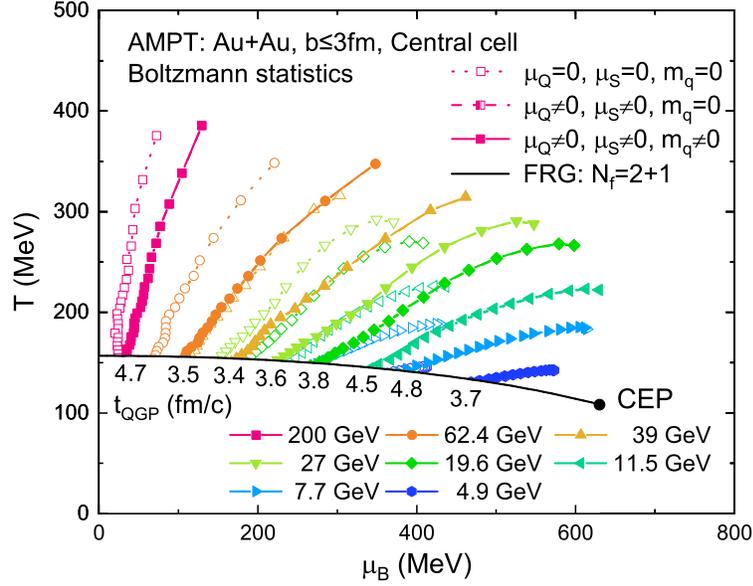}
	\caption{(Color	online) AMPT results on the average trajectories of the central cell in central Au$+$Au collisions at different energies in the QCD phase diagram. Three cases are compared: (I) $\mu_Q = 0$, $\mu_S = 0$, $m_q = 0$ (open symbols); (II) $\mu_Q \ne 0$, $\mu_S \ne 0$, $m_q = 0$ (half open symbols); (III) $\mu_Q \ne 0$, $\mu_S \ne 0$, $m_q \ne 0$ (filled symbols). The black curve shows the crossover phase boundary with the critical endpoint obtained from the functional renormalization group approach with $N_f = 2+1$ \cite{Fu:2019hdw}. The corresponding lifetime during which each trajectory stays in the QGP phase is also shown. 
	}
	\label{phase_diagram}
\end{figure}

In Fig.~\ref{phase_diagram}, we present the event-averaged evolution trajectory of the central cell of the partonic matter produced in central Au$+$Au collisions at different beam energies from the moment when the baryon chemical potential reaches the maximum value to the moment when it reaches the crossover curve in the QCD phase diagram of temperature and baryon chemical potential. Note that the crossover phase boundary is obtained from the functional renormalization group (FRG) method with $N_f = 2+1$, which agrees well with the phase boundary from the lattice QCD \cite{Fu:2019hdw}. From the filled symbols that represent the full consideration in which all chemical potentials and quark mass are included, we find that the partonic stage can last 3.4-4.8 fm/$c$ if the time when the system stays above the phase boundary is counted, which is consistent with the previous AMPT results for mid-central Au$+$Au collisions~\cite{Chen:2009cju}, but longer than the lifetime for the matter averaged over the transverse area from a semi-analytical calculation \cite{Mendenhall:2021maf}. If we take the location of the critical endpoint at ($T_{CEP}, \mu_{B_{CEP}}$) = (107, 635) MeV from the FRG calculation, the beam energies lower than 4.9 GeV~\cite{Fu:2019hdw, Andronic:2017pug} seem to be the most promising to reach the CEP, which could be accessed at fixed-target experiments at RHIC. Note that it has been found that the chemical and kinetic freeze-out parameters extracted from the AMPT model agree with the RHIC experimental measurements~\cite{Wang:2020wvu}. 

We further study the influences of the $\mu_Q$, $\mu_S$, and quark current mass $m_q$ on the event-averaged evolution trajectories (see Appendix~\ref{Boltzmann statistics}), as shown by half open and open symbols in Fig.~\ref{phase_diagram}. We can see that the influence of the quark mass is so small that the filled and half open symbols most overlap, because the current quark masses we use here are very small compared with the temperature and baryon chemical potential. However, we can observe that there is a large difference between filled or half open and open symbols, which indicates that $\mu_Q$ and $\mu_S$ are important for drive the evolution of the system.

\begin{figure}[htbp]
	\centering\includegraphics[scale=0.4]{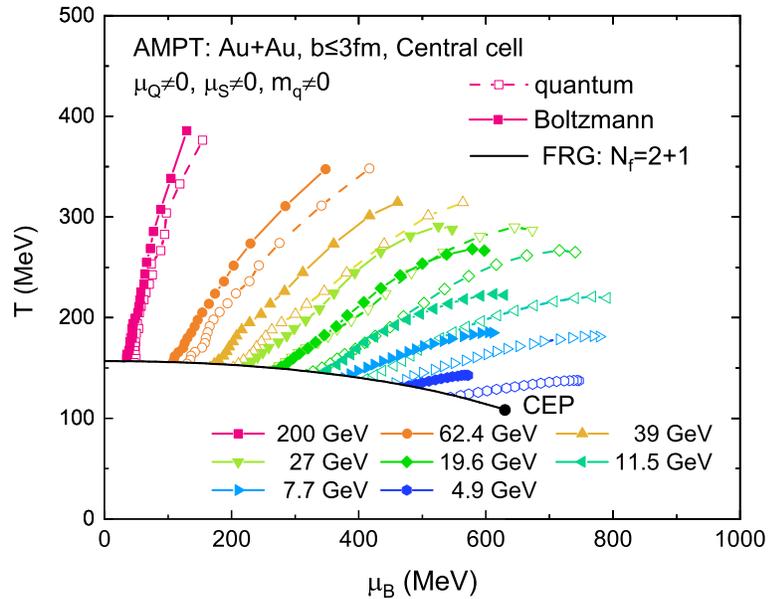}
	\caption{(Color	online) The average trajectory of the central cell in central Au$+$Au collisions at different energies in the QCD phase diagram of temperature versus baryon chemical potential from Boltzmann statistics (filled symbols) and the quantum statistics (open symbols). }
	\label{phase_diagram_statistics}
\end{figure}

Furthermore, we check whether the different statistics (see Appendices~\ref{Boltzmann statistics} and \ref{Quantum statistics}) can result in a difference of trajectories in the QCD phase diagram. We compare the results from Boltzmann statistics (filled symbols) and the quantum statistics (open symbols) in Fig.~\ref{phase_diagram_statistics}. We can see that with the decrease of collision energy, the difference between the two trajectories from the two statistics becomes larger. In general, a higher $\mu_B$ is obtained by the quantum statistics than that obtained by Boltzmann statistics, since the Pauli exclusion begins to play a role as $\mu_B$ increases, while this effect is absent in the Boltzmann statistics. Because the AMPT model assumes Boltzmann statistics, the results in the rest of this paper are presented using Boltzmann statistics.

\begin{figure}[h]
	\centering\includegraphics[scale=0.35]{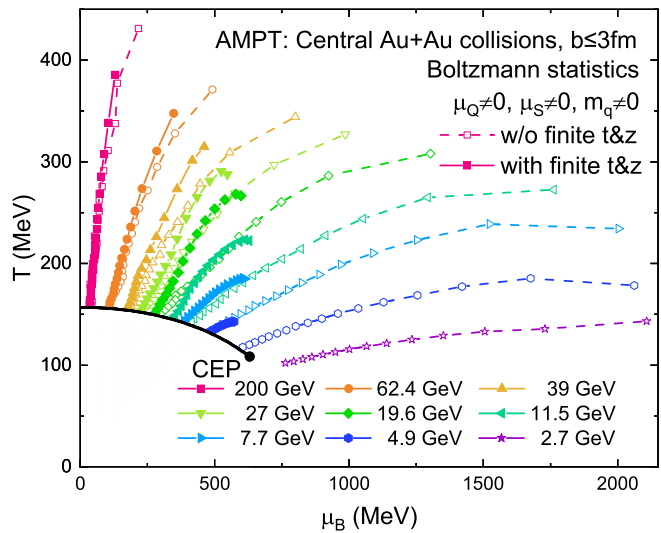}
	\caption{(Color	online) The average trajectory of the central cell in central Au$+$Au collisions at different energies in the QCD phase diagram from Boltzmann statistics with (filled symbols) and without (open symbols) including the finite nuclear thickness.}
	\label{phase_diagram_thickness}
\end{figure}

In addition, the finite thickness of nuclei is expected to affect the evolution trajectories in the QCD phase diagram, especially at low energies \cite{Lin:2017lcj, Mendenhall:2020fil, Mendenhall:2021maf}. In Fig.~\ref{phase_diagram_thickness} we compare the average trajectories with and without including the finite thickness for the central cell in central Au$+$Au collisions at different energies in the QCD phase diagram based on full consideration of Boltzmann statistics. We do not see any obvious change of evolution trajectory for the top RHIC energy, but the difference becomes more and more significant with the decrease of collision energy. For lower energies, the results without considering the finite-thickness effect start at much higher temperature and larger baryon chemical potential. For example, when considering the finite-thickness effect, the trajectory for 2.7 GeV disappears below the phase-transition boundary. Therefore, it is clearly necessary to properly include the finite nuclear thickness effect, especially for simulating heavy-ion collisions at low beam energies. 

\begin{figure}[htbp]
	\centering\includegraphics[scale=0.4]{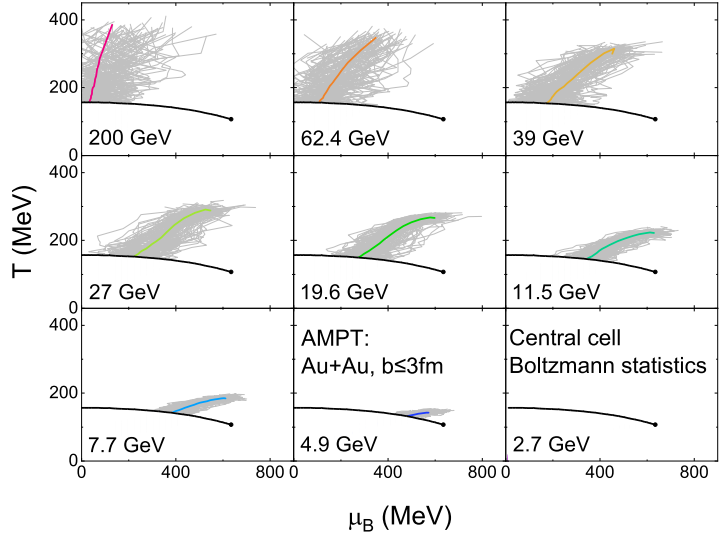}
	\caption{(Color	online) AMPT results on event-by-event trajectories of the central cell in central Au$+$Au collisions at different beam energies in the QCD phase diagram.}
	\label{byevent}
\end{figure}

We note that the above results come from the average of 10 000 central Au$+$Au events. However, event-by-event fluctuations can not be neglected, and these fluctuations could affect the search for the CEP of the QCD phase diagram. Figure.~\ref{byevent} shows the event-by-event trajectories of central Au$+$Au collisions at different beam energies from the AMPT-SM model. To suppress the effect from volume fluctuation, a multiplicity cut is further applied, in which we divide the total central events into 100 bins by multiplicity and only use the events in one middle bin around the average. Even so, we can see that the fluctuation of evolution trajectory is still large, especially at high energies, which could be due to larger volume fluctuations at higher energies.

It should be noted that the QGP created in high-energy heavy-ion collisions, which may consist of gluons and quarks in or near chemical and thermal equilibrium, should be governed by nonperturbative QCD interactions, which are missing in our model. Furthermore, the method that we used to extract temperature and baryon chemical potential only works for a noninteracting parton system in principle. Our extraction method assumes that all partons in the cell are in full thermal and chemical equilibrium~\cite{Lin:2014tya}; therefore, the extracted temperature and chemical potentials are the effective values if the system is in partial of thermal and/or chemical equilibrium. In addition, we focus on the central space-time rapidity and only study the partonic matter without the subsequent phase transition and hadronic evolution.  

\subsection{Equilibrium or nonequilibrium}
\label{Pressureanisotropy}
In the central cell of central Au$+$Au collisions, due to the cylindrical symmetry around the beam axis, the two transverse pressure components $P_{x}$ and $P_{y}$ are equal. Therefore, the transverse pressure can be defined to be $P_{T} = (P_{x}+P_{y})/2$~\cite{Zhang:2008zzk}, while the longitudinal pressure $P_{L}$ is just $P_{z}$. For a system in thermal equilibrium, its pressure must be isotropic, which satisfies the relation of $P_{T} = P_{L}=P$; otherwise, we
define the total pressure as $P = (P_{x}+P_{y}+P_{z})/3$. Therefore, a pressure anisotropy parameter, $P_{L}/P_{T}$, is defined to describe the degree of pressure anisotropy of the system. The closer the value of $P_{L}/P_{T}$ is to unity, the closer the system is to the state of thermal equilibrium.

\begin{figure}[htbp]
\centering\includegraphics[scale=0.4]{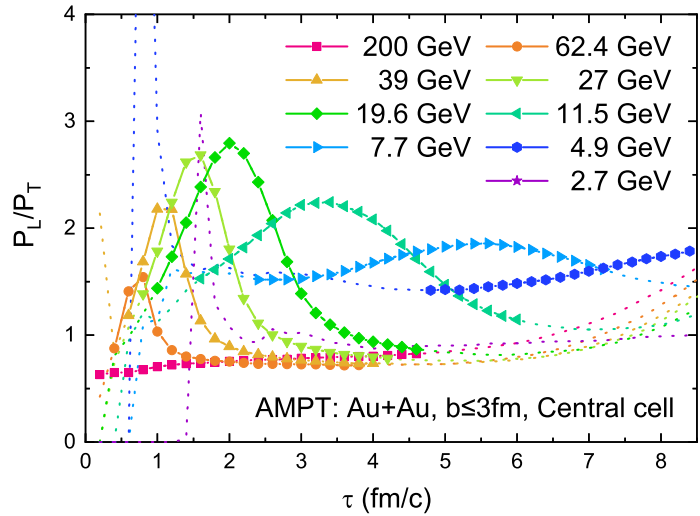}
\caption{(Color	online)  AMPT results for the time evolution of the pressure anisotropy parameter when its temperature and baryon chemical potential are above (filled symbols) and below (dotted curves) the phase boundary in the QCD phase diagram in the central cell in central Au$+$Au collisions at different beam energies.}
\label{PLPT}
\end{figure}

Figure.~\ref{PLPT} shows how the pressure anisotropy parameter in the central cell evolves with proper time in central Au$+$Au collisions at different beam energies. For Au$+$Au collisions at 200 GeV, we can see that $P_{L}/P_{T}$ keeps increasing, but still can not reach unity up to 5 fm/$c$. It indicates that even for the top RHIC energy, the central cell of the system actually does not reach thermal equilibrium when it arrives at the phase boundary in the AMPT model, which is consistent with previous results~\cite{Zhang:2008zzk, Lin:2014tya}. For lower energies, $P_{L}/P_{T}$ first increases up to a peak and decreases into a valley, and finally increases gradually due to the finite nuclear thickness. However, none of them reaches thermalization during the partonic stage. It shows that it is indeed different from the equilibrium evolution of hydrodynamical models.

\begin{figure}[htbp]
\centering\includegraphics[scale=0.4]{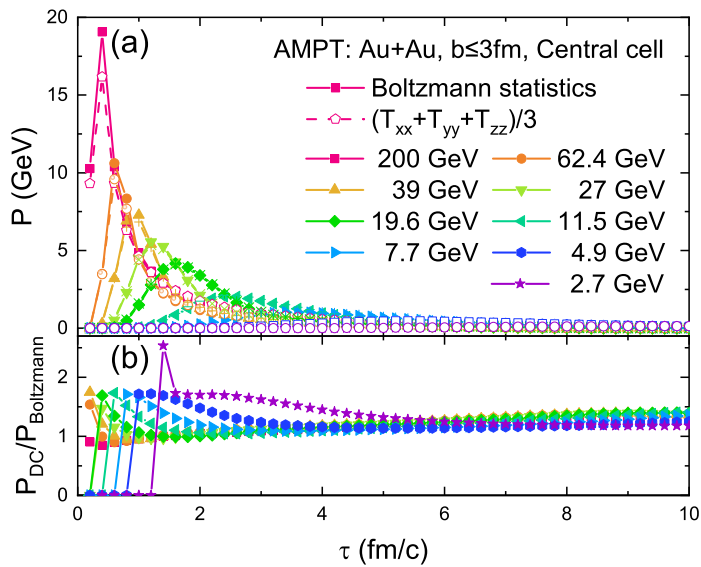}
\caption{(Color	online) AMPT results on the time evolution of (a) three diagonal components of the energy-momentum tensor ($P_{DC}$; open symbols) and the pressure from the Boltzmann statistical model ($P_{Boltzmann}$; filled symbols) in the central cell in central Au$+$Au energies, and (b) the ratio of $P_{DC}$ and $P_{Boltzmann}$.}
\label{Pressure}
\end{figure}

The total pressure can be extracted from the Boltzmann statistical model via:
    \begin{eqnarray}
    P(T)=\sum_i d_i \int \frac{d^3p}{(2\pi)^3}\frac{p^2}{3E_i(p, T)}f_B(p, T), 
    \label{p_Boltzmann}
    \end{eqnarray}
where $d_i$ is the degeneracy of partonic matter, $f_B(p, T)$ is the Boltzmann statistical distribution function, and $T$ is the temperature extracted from the Boltzmann statistical model. Figure.~\ref{Pressure} compares the pressure from three diagonal components of the energy-momentum tensor ($P_{DC}$) and from the Boltzmann statistical model ($P_{Boltzmann}$) in the central cell for central Au$+$Au energies. One can find that they are different especially for earlier time at lower energies, which indicates more extreme nonequilibrium of the system exists there. 

\begin{figure}[htbp]
\centering\includegraphics[scale=0.4]{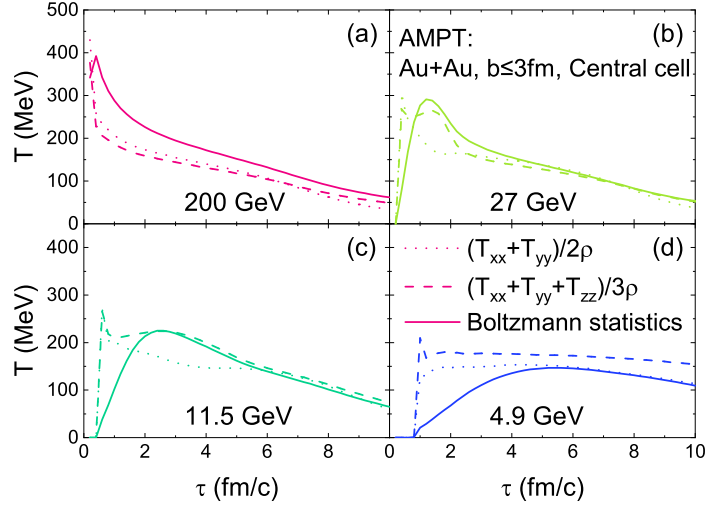}
\caption{(Color	online) AMPT results on the time evolution of the effective temperature extracted from transverse (dotted curves), three (dashed curves) diagonal components of the energy-momentum tensor, and the Boltzmann statistical model (solid curves) in the central cell in central Au$+$Au collisions at (a)200 GeV, (b) 27 GeV, (c) 11.5 GeV, and (d) 4.9 GeV.}
\label{T_comparison}
\end{figure}
The effective temperature can be defined locally by the ratio between the average of the diagonal components of the energy-momentum tensor and the density of all particles~\cite{Sorge:1995pw}. The effective temperature extracted from the diagonal components of the energy-momentum tensor and the Boltzmann statistical model in the central cell are shown in Fig.~\ref{T_comparison}. One can see that the effective temperatures extracted from the diagonal components of the energy-momentum tensor are different from our temperature especially for lower energies, although they give consistent trends. It is not only due to the nonequilibrium of the system, but also because our temperature extraction also considers the chemical potentials of conserved charges, especially the baryon chemical potential. In this sense, we should emphasize again that since the parton systems in Au$+$Au collisions at different energies from the AMPT model are not in complete equilibrium, the thermodynamic properties that we extracted above could be only approximate.

\section{Summary}
\label{summary}
We have studied the space-time evolution of the parton matter produced in central Au$+$Au collisions at different collision energies using the AMPT model with string melting and the finite nuclear thickness effect. The space-time evolutions of parton density and transverse flow is first presented for different collision energies. Then we extract the effective temperature and chemical potentials of the partons in the central cell based on Boltzmann statistics and quantum statistics. The temperature and baryon chemical potential first increase and then decrease with time, but their dependencies on the collision energy are opposite. By investigating the evolution of the partonic matter created in Au$+$Au collisions from 2.7 to 200 GeV, we obtain their evolution trajectories in the QCD phase diagram. The results indicate that the partonic state in the central cell exists for 3.4-4.8 fm/$c$ over this wide range of energies, and the trajectory depends on the statistics and whether the finite nuclear thickness is considered. We observe that the event-by-event trajectory fluctuates widely in the phase diagram. However, the evolution of pressure anisotropy indicates that only partial thermalization can be achieved when the partonic systems reach the predicted QCD phase boundary. Further studies of the evolution and the thermodynamic properties of the matter in heavy-ion collisions are indispensable for studying the QCD phase structure and the search for the critical point in experiments.

\begin{acknowledgments}
We thank Todd Mendenhall for checking the results in the Appendices. This work is supported in part by the National Natural Science Foundation of China under Grants No. 12147101, No. 11961131011, No. 11890710, No. 11890714, and No. 11835002, the Strategic Priority Research Program of Chinese Academy of Sciences under Grant No. XDB34030000,  and the Guangdong Major Project of Basic and Applied Basic Research under Grant No. 2020B0301030008 (H.-S. W. and G.-L.M.), the National Science Foundation under Grant No. PHY-2012947 (Z.-W.L.), and the National Natural Science Foundation of China under Contract No. 11775041 (W.-j. F.).

\end{acknowledgments}
\appendix
\section{Boltzmann statistics of partonic matter}
\label{Boltzmann statistics}
     In Boltzmann statistics, for the noninteracting thermal equilibrium system of  up, down, or strange quarks and gluon, the net parton number density and energy density are:
    \begin{eqnarray}\label{Boltzmann}
    n_{q}&=&\frac{N_c}{\pi^2} \int _{0}^{\infty} p^2 \Big [{e^{-(E(p) - \mu_{q})/T}} - e^{-(E(p) + \mu_{q})/T}\Big ]dp, \\
    \epsilon_{q}&=&\frac{N_c}{\pi^2} \int _{0}^{\infty} p^2 E(p) \Big [{e^{-(E(p) - \mu_{q})/T}} + e^{-(E(p) + \mu_{q})/T}\Big ]dp, \\
    n_{g}&=&\frac{N_c^2 - 1}{\pi^2} \int _{0}^{\infty} p^2 {e^{-E(p)/T}}dp,\\
    \epsilon_{g}&=&\frac{N_c^2 - 1}{\pi^2} \int _{0}^{\infty} p^2 E(p) {e^{-E(p)/T}}dp,
    \end{eqnarray}
    where $E(p)=(p^2+m_q^2)^{1/2}$ is the energy of parton, $T$ is the temperature, $N_c = 3$ is the number of colours, $n_{q}$, $\epsilon_{q}$, $n_{g}$, $\epsilon_{g}$, $m_{q}$, and $\mu_{q}$ are the net quark number density, the quark energy density, the gluon number density, the gluon energy density, the quark mass, and the quark chemical potential, respectively. $q$ represents the up, down, or strange quark. In this study, the quark masses are taken to be the current quark masses from the PDG~\cite{ParticleDataGroup:2014cgo}, e.g., $m_{u}$=2.16 MeV/$c^2$, $m_{d}$=4.67 MeV/$c^2$, and $m_{s}$=93 MeV/$c^2$.
    Relations among different chemical potentials are:
    \begin{eqnarray}
    \mu_u &=& \frac{1}{3} \mu_B + \frac{2}{3} \mu_Q,  \label{upotentials} \\
    \mu_d &=& \frac{1}{3} \mu_B - \frac{1}{3} \mu_Q,   \label{dpotentials} \\
    \mu_s &=& \frac{1}{3} \mu_B - \frac{1}{3} \mu_Q - \mu_S, 
    \label{spotentials}
    \end{eqnarray}
    where $\mu_B$, $\mu_Q$, and $\mu_S$ are the chemical potentials of baryon number, electric charge, and strangeness, respectively. Furthermore, the net conserved charge density and total energy density are
    \begin{eqnarray}\label{density}
    n_B &=& \frac{1}{3}(n_u + n_d + n_s),  \\
    n_Q &=& \frac{1}{3}(2 n_u - n_d - n_s),  \\
    n_S &=& - n_s, \\
    \epsilon_{total} &=& \epsilon_u + \epsilon_d + \epsilon_s + \epsilon_g,
    \end{eqnarray}
    We can obtain the four equations for conserved charge densities $n_{B}$, $n_{Q}$, and $n_{S}$ and $\epsilon_{total}$ as
    \begin{eqnarray}
    n_B &=& \frac{2}{\pi ^2} T \bigg[m_{d}^2 K_2\left(\frac{m_{d}}{T}\right) \sinh
   \left(\frac{\mu_{B}-\mu_{Q}}{3 T}\right)+m_{s}^2
   K_2\left(\frac{m_{s}}{T}\right) \sinh \left(\frac{\mu_{B}-\mu_{Q}-3
   \mu_{S}}{3 T}\right)  \notag \\
   & &+m_{u}^2 K_2\left(\frac{m_{u}}{T}\right) \sinh
   \left(\frac{\mu_{B}+2 \mu_{Q}}{3 T}\right)\bigg],  \notag \\
    n_Q &=& -\frac{2}{\pi
   ^2} T \bigg[m_{d}^2 K_2\left(\frac{m_{d}}{T}\right) \sinh \left(\frac{\mu_{B}-\mu_{Q}}{3
   T}\right)+m_{s}^2 K_2\left(\frac{m_{s}}{T}\right) \sinh \left(\frac{\mu_{B}-\mu_{Q}-3 \mu_{S}}{3
   T}\right)  \notag \\ 
   & &-2 m_{u}^2 K_2\left(\frac{m_{u}}{T}\right) \sinh \left(\frac{\mu_{B}+2 \mu_{Q}}{3 T}\right)\bigg],  \notag \\
    n_S &=& -\frac{6}{\pi ^2} m_{s}^2 T K_2\left(\frac{m_{s}}{T}\right) \sinh \left(\frac{\mu_{B}-\mu_{Q}-3 \mu_{S}}{3
   T}\right), \notag \\
    \epsilon_{total} &=& \frac{6}{\pi ^2} T \bigg[8 T^3+m_{u}^3
   K_1\left(\frac{m_{u}}{T}\right) \cosh \left(\frac{\mu_{B}+2 \mu_{Q}}{3 T}\right)+3 m_{u}^2 T K_2\left(\frac{m_{u}
   }{T}\right) \cosh \left(\frac{\mu_{B}+2 \mu_{Q}}{3 T}\right)  \notag \\
   & &+m_{s}^3
   K_1\left(\frac{m_{s}}{T}\right) \cosh \left(\frac{\mu_{B}-\mu_{Q}-3 \mu_{S}}{3 T}\right)+3 m_{s}^2
   T K_2\left(\frac{m_{s}}{T}\right) \cosh \left(\frac{\mu_{B}-\mu_{Q}-3 \mu_{S}}{3 T}\right)  \notag \\
   & &+m_{d}^3 K_1\left(\frac{m_{d}}{T}\right) \cosh \left(\frac{\mu_{B}-\mu_{Q}}{3 T}\right)+3
   m_{d}^2 T K_2\left(\frac{m_{d}}{T}\right) \cosh \left(\frac{\mu_{B}-\mu_{Q}}{3 T}\right)\bigg],
    \label{mu_T} 
    \end{eqnarray}
    where $K_1$ and $K_2$ are Bessel functions of the second kind.
  Specially, if $m_q$ is neglected, Eqs.~(\ref{mu_T}) can be simplified as:
    \begin{eqnarray}
    n_B &=& \frac{4  T^3}{\pi ^2} \sinh \left(\frac{\mu_B-\mu_Q-3 \mu_S}{3 T}\right)+\frac{4  T^3
   }{\pi ^2}\sinh \left(\frac{\mu_B-\mu_Q}{3 T}\right)  \notag \\
    & &+\frac{4  T^3 }{\pi ^2}\sinh \left(\frac{\mu_B+2
   \mu_Q}{3 T}\right),  \notag \\
    n_Q &=& -\frac{4  T^3 }{\pi
   ^2}\left(\sinh \left(\frac{\mu_B-\mu_Q-3 \mu_S}{3 T}\right)+\sinh
   \left(\frac{\mu_B-\mu_Q}{3 T}\right)-2 \sinh \left(\frac{\mu_B+2 \mu_Q}{3 T}\right)\right),  \notag \\
    n_S &=& -\frac{12 T^3 }{\pi ^2}\sinh \left(\frac{\mu_B-\mu_Q-3 \mu_S}{3 T}\right), \notag \\
    \epsilon_{total} &=& \frac{12 T^4}{\pi ^2}\bigg[3 \cosh \left(\frac{\mu_B-\mu_Q-3 \mu_S}{3 T}\right)+3 \cosh
   \left(\frac{\mu_B-\mu_Q}{3 T}\right)  \notag \\
    & &+3 \cosh \left(\frac{\mu_B+2 \mu_Q}{3
   T}\right)+4\bigg].
    \label{Boltzmann_mu_T_simpl1} 
    \end{eqnarray}
    Furthermore, if $n_{Q}$ and $n_{S}$ are neglected, Eqs.~(\ref{Boltzmann_mu_T_simpl1}) can be simplified as:
    \begin{eqnarray}
    n_B &=& \frac{12 T^3 \sinh\left [\mu_B/(3 T)\right ]}{\pi^2},  \notag \\
    n_Q &=& 0,  \notag \\
    n_S &=& -\frac{12 T^3 \sinh\left [\mu_B/(3 T)\right ]}{\pi^2}, \notag \\
    \epsilon_{total} &=& \frac{48 T^4}{\pi^2}+\frac{108 T^4 \cosh\left [\mu_B/(3 T)\right ]}{\pi^2}.
    \label{Boltzmann_mu_T_simpl2} 
    \end{eqnarray}

\section{Quantum statistics of partonic matter}
\label{Quantum statistics}
    In quantum statistics, the net parton number and energy densities are:
    \begin{eqnarray}\label{Quantum}
    n_{q}&=&\frac{N_c}{\pi^2} \int _{0}^{\infty} p^2 \Big [\frac{1}{e^{(E(p) - \mu_{q})/T}+1} - \frac{1}{e^{(E(p) + \mu_{q})/T}+1}\Big ]dp, \\
    \epsilon_{q}&=&\frac{N_c}{\pi^2} \int _{0}^{\infty} p^2 E(p) \Big [\frac{1}{e^{(E(p) - \mu_{q})/T}+1} + \frac{1}{e^{(E(p) + \mu_{q})/T}+1}\Big ]dp, \\
    n_{g}&=&\frac{N_c^2 - 1}{\pi^2} \int _{0}^{\infty} \frac{p^2 dp}{e^{E(p)/T}-1},\\    
    \epsilon_{g}&=&\frac{N_c^2 - 1}{\pi^2} \int _{0}^{\infty} \frac{p^2 E(p) dp}{e^{E(p)/T}-1}.
    \end{eqnarray}
 Similar to Boltzmann statistics, using the relations ((\ref{upotentials})-(\ref{dpotentials}), and \ref{spotentials}), we can also get four equations about $n_{B}$, $n_{Q}$, $n_{S}$, and $T$ under quantum statistics. Specially, if $m_q$ is neglected, the four equations can be simplified as:
     \begin{eqnarray}
    n_B &=& \frac{1}{27 \pi ^2}\mu_B^3-\frac{1}{9 \pi ^2}\mu_B^2 \mu_S+\frac{ 1
   }{9 \pi 
   ^2}\mu_B\left[3 \left(\pi ^2 T^2+\mu_S^2\right)+2 \mu_Q \mu_S+2 \mu_Q^2\right]  \notag \\
    & &-\frac{1}{27 \pi ^2} \left(9\pi ^2 \mu_S T^2-2 \mu_Q^3+3 \mu_Q^2 \mu_S+9 \mu_Q \mu_S^2+9\mu_S^3\right),  \notag \\
    n_Q &=& \frac{ 1}{9 \pi
   ^2}\mu_Q\left(6 \pi ^2 T^2+2 \mu_B^2-2 \mu_B \mu_S+3 \mu_S^2\right)  \notag \\
    & &+\frac{1 }{9 \pi ^2}\mu_S\left(3\pi ^2
   T^2+\mu_B^2-3 \mu_B \mu_S+3\mu_S^2\right) +\frac{ 1 }{9 \pi ^2}\mu_Q^2(2 \mu_B+\mu_S)+\frac{2
   }{9 \pi ^2}\mu_Q^3,  \notag \\
    n_S &=& -\frac{1}{27 \pi ^2}\mu_B^3 +\frac{1 }{9 \pi
   ^2}\mu_B^2(\mu_Q+3 \mu_S) -\frac{1}{9 \pi
   ^2}\mu_B \left(3 \pi ^2 T^2+\mu_Q^2+6 \mu_Q \mu_S+9 \mu_S^2\right)   \notag \\
    & &+\frac{1}{27 \pi ^2} (\mu_Q+3 \mu_S) \left(9\pi
   ^2 T^2+\mu_Q^2+6 \mu_Q \mu_S+9 \mu_S^2\right), \notag \\
    \epsilon_{total} &=& \frac{1}{36 \pi ^2}\mu_B^4 -\frac{1}{9 \pi ^2}\mu_B^3 \mu_S +\frac{
   1}{6 \pi
   ^2}\mu_B^2 \left(3\pi ^2 T^2+2 \mu_Q^2+2 \mu_Q \mu_S+3 \mu_S^2\right)  \notag \\
    & &-\frac{1 }{9 \pi ^2}\mu_B\left(9\pi ^2 \mu_S T^2-2 \mu_Q^3+3 \mu_Q^2 \mu_S+9 \mu_Q \mu_S^2+9
   \mu_S^3\right)   \notag \\
    & &+\frac{1}{36 \pi ^2}[3\pi ^2 T^2 (19 \pi ^2 T^2+12\mu_Q^2 +12\mu_Q \mu_S
   +18 \mu_S^2 ) \notag \\
    & &+6 \mu_Q^4 +4 \mu_Q^3
   \mu_S +18 \mu_Q^2  \mu_S^2  +27 \mu_S^4 +36 \mu_Q \mu_S^3].
    \label{quantum_mu_T_simpl1} 
    \end{eqnarray}
  Furthermore, if $n_{Q}$, $n_{S}$, and $m_q$ are neglected, Eqs.~(\ref{quantum_mu_T_simpl1}) can be simplified as:
    \begin{eqnarray}
    n_B &=& \frac{\mu_B^3}{27 \pi^2} + \frac{\mu_B T^2}{3},  \notag \\
    n_Q &=& 0,  \notag \\
    n_S &=& -\frac{\mu_B^3}{27 \pi^2} - \frac{\mu_B T^2}{3}, \notag \\
    \epsilon_{total} &=& \frac{\mu_B^4}{36 \pi^2} + \frac{\mu_B^2 T^2}{2} + \frac{19\pi^2 T^4}{12}.
    \label{quantum_mu_T_simpl2} 
    \end{eqnarray}

\end{document}